\documentstyle[12pt]{article}

\begin{document}

\thispagestyle{empty}
\def\pubnum{343}
\def\data{September, 1994}

\begin{flushright}
{\parbox{3.5cm}{
UAB-FT-343

September, 1994

hep-ph/9410310
}}
\end{flushright}

\vspace{1cm}
\hyphenation{ne-ver-the-less}
\hyphenation{cano-nical ca-nonical canoni-cal}
\hyphenation{ob-ser-va-ble}
\begin{center}
\begin{large}
\begin{bf}
SUPERSYMMETRIC ELECTROWEAK RENORMALIZATION OF THE $Z$-WIDTH IN THE MSSM (I)\\

\end{bf}

\end{large}
\vspace{1cm}
David GARCIA
\footnote{Internet address:GARCIA@IFAE.ES}\,,
Ricardo A. JIM\'ENEZ
\footnote{Internet address:ABRAHAM@IFAE.ES}\,,
Joan SOL\`A
\footnote{Internet addresses:IFTESOL@CC.UAB.ES and SOLA@IFAE.ES}

\vspace{0.25cm}
Grup de F\'{\i}sica Te\`orica\\

and\\

Institut de F\'\i sica d'Altes Energies\\

\vspace{0.25cm}
Universitat Aut\`onoma de Barcelona\\
08193 Bellaterra (Barcelona), Catalonia, Spain\\
\end{center}
\vspace{0.3cm}
\begin{center}
{\bf ABSTRACT}
\end{center}
\begin{quotation}
\noindent
\hyphenation{ana-ly-ses a-na-ly-ses}
\hyphenation{de-li-ca-te}
\hyphenation{phe-no-me-no-lo-gi-cal}

Within the framework of the MSSM, we compute the complete set of electroweak
one-loop supersymmetric quantum effects on the width $\Gamma_Z$ of the
$Z$-boson in the on-shell renormalization scheme.
Numerical analyses of the corrections to the various partial widths into
leptons and quarks are presented. On general grounds,
the average size of the electroweak SUSY corrections to $\Gamma_Z$ may
well saturate the level of the present theoretical uncertainties,
even if considering the full supersymmetric spectrum
lying in the neighbourhood of
the unaccessible LEP 200 range. Remarkably enough,
for the present values of the top quark mass,
the electroweak SUSY effects could be, globally, very close or even bigger
than the electroweak SM corrections, but
opposite in sign. Therefore, in the absence of theoretical errors, there
are large regions of parameter space where one
could find that, effectively,
the electroweak SM corrections are ``missing'', or even having the ``wrong''
sign.
This should be helpful in discriminating between the SM and the MSSM.
However, an accurate prediction
of the electroweak quantum effects on $\Gamma_Z$ will only
be possible, if  $\Delta r$ and $\alpha_s$ are
pinned down in the future with enough precision.
\end{quotation}

\baselineskip=6.5mm

\newpage

An outstanding laboratory to test possible manifestations of Supersymmetry
(SUSY)
and particularly of the Minimal Supersymmetric Standard Model
(MSSM)\,\cite{b1001}
is LEP, either at $Z$-pole energies
or in the near future also from on-shell $W$-physics.
If no direct production of supersymmetric particles (``sparticles'')
is achieved neither at LEP 100 nor at LEP 200, still some
indirect manifestations could be discovered from quantum effects
\,\cite{b1002}.
In fact, radiative corrections to conventional
physical processes\,\cite{b1003}
are a powerful tool to search for mass scales within and
beyond the Standard Model (SM), and they offer us the opportunity
to peep at sectors of the
theory that are not (yet) directly observable.
In this respect it is useful to remember that
at LEP 200 the $W$-mass will be measured with a remarkable precision of
$\delta M_W=\pm 28\,(stat.)\pm 24\,(syst.)\, MeV$\,\cite{b1004}. Recent
analyses\,
\cite{b1005,b1006} have shown that a measurement of the $W$-mass
with that precision, or even a factor of two worse, would enable us to hint at
virtual SUSY effects even if the full supersymmetric spectrum lies in the
vicinity of the unaccessible LEP 200 range
(\,$\stackrel{\scriptstyle >}{{ }_{\sim}} 100\,GeV$).

Similarly, on-shell $Z$-physics is also sensitive to
quantum effects from sparticles.
The mass and width of the $Z$-boson are experimentally known with particularly
good accuracy\,\cite{b1007}
\begin{equation}
M_Z^{\rm exp}=91.1895\pm 0.0044\,GeV
\label{eq:MZexp}
\end{equation}
and
\begin{equation}
\Gamma_Z^{\rm exp}=2.4969\pm 0.0038\,GeV\,.
\label{eq:GZexp}
\end{equation}
One expects that the $Z$-width will eventually be measured within only
 $|\delta\Gamma^{\rm exp}_Z|\stackrel{\scriptstyle <}{{ }_{\sim}}
2\,MeV$\,\cite{b1007}.
In the on-shell scheme\,\cite{b10B8}
the $Z$-mass enters as an experimental input,
while the $Z$-width
can be predicted with great accuracy. A detailed calculation of the electroweak
one-loop effects on $\Gamma_Z$ in the SM
is given in refs.\cite{b1009,b1010}.
The final updated numerical result, including QCD corrections,
reads\,\cite{b1011}
\begin{equation}
\Gamma_Z^{SM}=2.4922\pm 0.0075\pm 0.0033\, GeV\,,
\label{eq:GZMSM1}
\end{equation}
where the label SM stresses that this theoretical
result drops from the strict
Standard Model with minimal (single) Higgs sector.
The first error corresponds to the variation with the top quark and Higgs mass
within the allowed range of $\delta M_W$ and $\Delta r$\,\cite{b1011},
and the second error is the hadronic
uncertainty from $\alpha_s=0.123\pm 0.006$
measured from hadronic event topologies
at the $Z$-peak. Inclusion of the recently claimed CDF value for the top quark
mass
( $m_t=174\pm 10\, ^{+13}_{-12}\,GeV$\,\cite{b1TOP}) leads to\,\cite{b1011}
\begin{equation}
\Gamma_Z^{SM}|_{CDF}=2.4933\pm 0.0064\pm 0.0033\, GeV\,.
\label{eq:GZMSM2}
\end{equation}
In spite of how respectably well the SM prediction
(\ref{eq:GZMSM1})-(\ref{eq:GZMSM2}) matches the
experimental result (\ref{eq:GZexp}), the inherent
theoretical and experimental uncertainties still
leave room enough to allocate hypothetical new
contributions beyond the SM, such as
those from the MSSM.
Indeed, in the MSSM we expect a different theoretical prediction for the $Z$
width,
$\Gamma_Z^{MSSM}$, which may be conveniently split up into two pieces
\begin{equation}
\Gamma_Z^{MSSM}=\Gamma_Z^{RSM}+\delta\Gamma_Z^{MSSM}\,,
\label{eq:GZMSSM}
\end{equation}
where $\Gamma_Z^{RSM}$ involves the contribution from a so-called
``Reference SM''(RSM)\,\cite{b1HAB}: namely, the Standard Model with the single
Higgs
mass set equal to the mass of the lightest $CP$-even Higgs scalar of the MSSM,
whereas  $\delta\Gamma_Z^{MSSM}$
constitutes the total quantum departure of the MSSM prediction
with respect to that Reference Standard Model.
Besides, $\delta\Gamma_Z^{MSSM}$ itself
splits up naturally into two parts, viz.
the extra two-doublet Higgs contribution $\delta\Gamma_Z^{H}$, in which
the single Higgs part included in $\Gamma_Z^{RSM}$ has been subtracted out
in order to avoid
double-counting, and the SUSY contribution $\delta\Gamma_Z^{SUSY}$
{}from the plethora of ``genuine'' ($R$-odd) supersymmetric particles:
\begin{equation}
\delta\Gamma_Z^{MSSM}=\delta\Gamma_Z^{H}+\delta\Gamma_Z^{SUSY}\,.
\label{eq:Tshift}
\end{equation}
 Difficult enough,
the new effects have to be disentangled from
the yield of conventional sectors
of the theory which are not yet experimentally determined with enough
precision,
most conspicuously the
top quark mass
\footnote{Potentially significant virtual
hints from SUSY have been recently recognized
on the physics of top quark decay\,\cite{b1B11}.}.
Furthermore, as it
has been pointed out in other contexts\,\cite{b1002}, the quantum corrections
in the MSSM, quite independently from the values of the
various parameters, may mimic those in the SM.
Therefore, to effectively discriminate the theoretical predictions
of the MSSM from those of
the SM is a rather delicate matter that requires to compare the simultaneous
predictions on several observables, as for
example $\Gamma(Z\rightarrow f\bar{f})$ and $M_W$. However, whereas the full
treatment of the SUSY
corrections to the $W$-mass in the on-shell scheme
 has already been accomplished in detail by several groups
\,\cite{b1002,b1005,b1006},
the corresponding corrections to $\Gamma_Z$ have
been partially computed\,\cite{b1002,b1012}
on only some specific decay channels and/or explicitly
ignoring the effects from parts of the SUSY spectrum and/or considering only
leading
effects (e.g. large Yukawa couplings).
Particularly interesting by itself is the study of the additional
contributions to  $\Gamma_Z$ from two-Higgs-doublet-model
extensions of the SM \,\cite{b1013}.
Although there is some work in the literature for the general unconstrained
case
and for the supersymmetric case  \,\cite{b1113,b1213}, a systematic analysis of
the latter incorporating  a detailed treatment of the mass
relations in the MSSM Higgs sector is lacking.
Therefore, in this note we would like to settle down these matters on the basis
of an exact one-loop calculation
of the electroweak part of $\delta\Gamma_Z^{MSSM}$
by keeping all effects from gauge and
Yukawa couplings and for arbitrary values of the parameters.
The calculation of this quantity is indeed a rather complex
task. We believe, also an important task. We have faced it in full, with
the double purpose of completing previous calculations and at the same time
to assess the real possibilities of SUSY to give a hint of existence
{}from the high precision world of $Z$ physics.
To our knowledge, no systematic analysis
of the impact of the electroweak genuine SUSY sector of the MSSM on
$\Gamma_Z$, split up into the various partial widths,
has been clearly put in a nutshell anywhere in the literature.
Our intention is to numerically demonstrate that, despite of the fact that
all SUSY effects must decouple for large enough sparticle masses, we may still
expect
potentially measurable supersymmetric electroweak
contributions (i.e. contributions that could be near the present theoretical
errors
and well above the
planned experimental accuracy
 $|\delta\Gamma^{\rm exp}_Z|\stackrel{\scriptstyle <}{{ }_{\sim}} 2\,MeV$)
even for an average SUSY spectrum that surpasses the  LEP 200 discovery range.
These corrections have to be added to possible
SUSY-QCD corrections \,\cite{b1QCD} mediated by gluinos,
which are generally smaller and of the
same sign.
On the whole, this should help to untangle the differences between the SM and
the
MSSM, especially when combined with the predictions on other observables like
$M_W$ and the asymmetries at the $Z$-pole.
We divide our presentation into two parts: i) In the present part (Part I)
a full account of the ``genuine'' ($R$-odd) SUSY contributions;
namely, from sfermions (squarks and sleptons) and
``inos'' (charginos and neutralinos), is considered in detail. They constitute
the complete supersymmetric electroweak radiative shift $\delta\Gamma_Z^{SUSY}$
in eq.(\ref{eq:Tshift}) and we find that, globally,
they could provide a source of relatively large loop contributions, in
particular
if the sparticles are not too heavy.
For definiteness,
our numerical analysis follows the same pattern of sparticle masses as defined
by
the so-called Models I and II in Ref.\cite{b1005}, where we analyzed the full
contribution to $\Delta r^{SUSY}$. These models are general enough
to comprehend both phenomenological as well as more restricted (supergravity
inspired\,\cite{b1001}) models.
On the other hand, in Part II\,\cite{b1PII},
which we present in a separate note following
this one, we consider the analysis of  $\delta\Gamma_Z^{H}$ within the
framework
of an improved (one-loop corrected\,\cite{b1313})
MSSM Higgs sector and compare
with $\delta\Gamma_Z^{SUSY}$. The result is especially significant for the
$b\bar{b}$
channel and the associated ratio $R_b$, since its experimental value
could be in discrepancy with the SM prediction\,\cite{b1007}.
However, we postpone the explicit presentation of our
analysis of the full width (\ref{eq:GZMSSM}) of the $Z$ in
the Minimal Supersymmetric Standard Model for a separate and
lengthy forthcoming publication where the gory details of
the present calculation can be found, together with a more comprehensive
exposition of the numerical results\,\cite{b1014}. In the meanwhile, our
notation
for the SUSY formalism follows
Ref.\cite{b1B11} and also the early work of Ref.\cite{b2003}.


As stated, our computation of $\delta\Gamma_Z^{SUSY}$ is carried out in the
on-shell renormalization scheme
\footnote{For a detailed review, see e.g.
refs.\cite{b1003,b2001,b2002}.}, where the fine structure constant,
 $\alpha\equiv\alpha_{\rm em} (q^2=0)$,
and the physical masses of the gauge bosons, fermions and scalars are the
renormalized
parameters:\,\, $(\alpha, M_W, M_Z, M_H, m_f,...)$. We will, for brevity
sake, refer to it as the $\alpha$-scheme: $(\alpha, M_W, M_Z)$.
In practice, in order to achieve higher accuracy in the theoretical predictions
it is convenient to adopt the constrained $\alpha$-scheme $(\alpha, G_F, M_Z)$
in which one substitutes the high precision effective parameter
$G_F$ (Fermi's constant in $\mu$-decay)
for $M_W$ by means of the constraint
\begin{equation}
{G_F\over\sqrt{2}}
={\pi\alpha\over{ 2\,M_Z^2\,s^2\,c^2} }\,{1\over 1-\Delta r^{MSSM}}
\label{eq:constraint}\,,
\end{equation}
where $c\equiv M_W/M_Z$ and $s^2\equiv 1-c^2$. In this equation,
\begin{equation}
\Delta r^{MSSM}=\Delta r(\alpha, M_W, M_Z, M_H, m_f, m_{SUSY},...)\,,
\end{equation}
where $m_{SUSY}$ is a generic soft SUSY-breaking parameter which sets the
characteristic mass scale of the various sparticles. $\Delta r^{MSSM}$
embodies the resultant finite effect from
all possible radiative corrections, universal (U)
and non-universal (NU) to $\mu$-decay in the MSSM:
\begin{equation}
 \Delta r^{MSSM}  = \Delta r^U+\Delta r^{NU}=
-{\hat{\Sigma}_W(0)\over M_W^2}+\Delta r^{NU}\,.
\label{eq:deltar1}
\end{equation}
For our purposes, it will also be useful to split $ \Delta r^{MSSM}$
as follows
\begin{equation}
\Delta r^{MSSM}=  \Delta r^{RSM}+\delta(\Delta r)^{MSSM}\,,
\label{eq:deltar2}
\end{equation}
with
\begin{equation}
 \delta (\Delta r)^{MSSM}=\Delta r^{H}+\Delta r^{SUSY}\,,
\label{eq:dDeltar}
\end{equation}
the meaning of the terms on the RHS of
eqs.(\ref{eq:deltar2})-(\ref{eq:dDeltar})
being fully parallel to those in
eqs.(\ref{eq:GZMSSM})-(\ref{eq:Tshift}).
The renormalized self-energy of the $W$-boson at zero frequency
is given by
\begin{equation}
\hat{\Sigma}_W (0) ={\Sigma}_W(0)+\delta M_W^2+M_W^2\left\{
\left.  {\Sigma^{\gamma}(k^2)\over k^2}\right|_{k^2=0}-2{c\over s}\,
{\Sigma^{\gamma Z}(0)\over M_Z^2}
+{c^2\over s^2}
({\delta M_Z^2\over M_Z^2}-{\delta M_W^2\over M_W^2})\right\}\,.
\label{eq:RSEW}
\end{equation}
Here $\Sigma^{\gamma, W, Z,...}(k^2)$ are the real parts of the
(transverse components of the unrenormalized) gauge boson self-energy
functions.
Finally, the gauge boson mass counterterms
\begin{equation}
\delta M_W^2=-\Sigma^W(k^2=M_W^2)\,\,,\ \ \ \ \
\delta M_Z^2=-\Sigma^Z(k^2=M_Z^2)
\label{eq:MCT}
\end{equation}
are enforced by the on-shell renormalization conditions.
{}From these equations one usually decomposes
\,\cite{b1003}
\begin{equation}
\Delta r^U= \Delta\alpha-{c^2\over s^2}\,\Delta\rho +\Delta r_{rem.}\,,
\label{eq:DeltarRG}
\end{equation}
whose interpretation in terms of the renormalization group (RG) running of
$\alpha$ ($\Delta\alpha$) and the various
statical ($\Delta\rho$) and dynamical ($\Delta r_{rem.}$) contributions to the
breaking of global $SU(2)$ symmetries (such as custodial
symmetry\,\cite{b2022})
is well known in the literature\,\cite{b2032}.
As for the explicit
analytic expressions of the SUSY contributions to the above formulas
we use the results of Ref.\cite{b2003}, which we shall not repeat here
\footnote{Although the (low-energy)
renormalization framework of Ref.\cite{b2003}
is different from the one considered here,
the unrenormalized expressions are the same and
they have been straightforwardly adapted ( as in Ref.\cite{b1005})
to the counterterm structure
of the minimal $\alpha$-scheme of Ref.\cite{b2001} without
changing the sign convention for the self-energy functions.}.

The one-loop partial width of the $Z$-boson into a fermion-antifermion
pair can be expressed generically
in the vector-axial representation, and in the $G_F$-parametrization,
in terms of two form factors
$\rho_f=1+\delta\rho_f$
and $\kappa_f=1+\delta\kappa_f$ as follows:
\begin{equation}
\Gamma (Z\rightarrow f\bar{f}) = N_c^f\,{G_F\,M_Z^3\over 24\,\pi\sqrt{2}}\,
\rho_f\,\sqrt{1-4\,\mu_f}\,
\left[1-4\,\mu_f+(1-4\,|Q_f|\,s^2\,\kappa_f)^2\,(1+2\,\mu_f)\right]\,,
\label{eq:Zwidth1}
\end{equation}
where
\begin{equation}
\rho_f={1-\Delta r^{MSSM}\over
 1-\hat{\Sigma}_Z'(M_Z^2)}+{2\over a_f}\,\delta a_f\,,
\label{eq:rhof}
\end{equation}
\begin{equation}
\kappa_f=1-{1\over 4\,|Q_f|\,s^2\,a_f}\,\left(\delta v_f
-{v_f\over a_f}\,\delta a_f\right)\,,
\label{eq:kappaf}
\end{equation}
and
\begin{equation}
s^2={1\over 2}\,\left\{1-\left[1-
{A\over (1-\Delta r^{MSSM})}\right]^{1/2}\right\}
\ \ \ \ \ (A\equiv {4\pi\alpha\over \sqrt{2}\,G_F\,M_Z^2})\,.
\label{eq:s2}
\end{equation}
Here $N_c^f=1$ (for leptons), $3$ (for quarks), $\mu_f\equiv m_f^2/M_Z^2$;
$v_f=(T_3^f-2\,Q^f\,s^2)/2sc$ and $a_f=T_3^f/2sc$
are the vector and axial coefficients of the neutral current
\,\cite{b1009}. The one-loop corrections to these
coefficients are defined through the radiative shifts
\begin{equation}
 v_f\rightarrow v_f+\delta v_f\ \ \,,\ \ a_f\rightarrow a_f+\delta a_f\,.
\end{equation}
These shifts, together with $\Delta r$, are to be computed in the MSSM
and isolated their departure from
the total ``background'' contribution
of the RSM (see eqs.(\ref{eq:Tshift}) and (\ref{eq:dDeltar})).
Similarly for the  $Z$ wave-function renormalization effects, which
are represented in eq.(\ref{eq:rhof}) by the derivative of the
corresponding renormalized self-energy :
\begin{equation}
\hat{\Sigma}_Z'(M_Z^2)= \Sigma_Z'(M_Z^2)-
\left.{\Sigma^{\gamma}(k^2)\over k^2}\right|_{k^2=0}+2{c^2-s^2
\over s\,c}\,{\Sigma^{\gamma Z}(0)\over M_Z^2}-{c^2-s^2\over s^2}
({\delta M_Z^2\over M_Z^2}-{\delta M_W^2\over M_W^2})\,.
\label{eq:RSEPZ}
\end{equation}
As for the renormalized $\gamma Z$ mixed self-energy on the mass shell of the
$Z$,
it is included as a part of the total radiative
shift of the vector coefficient as follows:
\begin{equation}
\delta v_f^{\gamma Z}=Q_f\,{\hat{\Sigma}^{\gamma Z} (M_Z^2)\over M_Z^2}
=Q_f\left\{{\Sigma^{\gamma Z}(M_Z^2)+\Sigma^{\gamma Z}(0)\over M_Z^2}-{c\over
s}
({\delta M_Z^2\over M_Z^2}-{\delta M_W^2\over M_W^2})\right\}\,.
\label{eq:RSGZ}
\end{equation}
We remark that the total {\sl additional} MSSM
contribution from $\Sigma^{\gamma Z}(k^2)$
with respect to the RSM turns out to vanish at $k^2=0$, and so $\Sigma^{\gamma
Z}(0)$
may actually be dropped from eqs.(\ref{eq:RSEW}), (\ref{eq:RSEPZ}) and
(\ref{eq:RSGZ}) .
The list of the Feynman diagrams and analytical MSSM
contributions from vertices and self-energies to the form
factors $\rho_f$ and $\kappa_f$ is provided in Ref.\cite{b1014} and,
as asserted, we shall not dwell here into their detailed structure.
Expanding up to one-loop level the previous formulae,
the general form of the SUSY correction to any given partial width
can be written (in a notation partly inspired from Ref.\cite{b1012})
\begin{equation}
\delta\Gamma^{SUSY}(Z\rightarrow f\bar{f})
=\Gamma_Z^0(G_F)\left[\,\nabla_U^{SUSY}+
\nabla_{Q_f}^{SUSY}+\nabla_{V_f}^{SUSY}\right]\,,
\label{eq:DeltaSUSY}
\end{equation}
where $\Gamma_Z^0(G_F)$ is the tree-level width in the $G_F$-parametrization
 (defined by eq.(\ref{eq:Zwidth1}) with
$\rho_f=\kappa_f=1$), and
\begin{eqnarray}
\nabla_U^{SUSY} &=&\hat{\Sigma}_Z'(M_Z^2)-\Delta r^{SUSY}=\Delta\rho^{SUSY}+...
\nonumber\\
\nabla_{Q_f}^{SUSY} &=& 2|Q_f|{v_f\,a_f\over v_f^2+a_f^2}
(4sc\,{\hat{\Sigma}^{\gamma Z}(M_Z^2)\over M_Z^2}-
{A\over \sqrt{1-A}}\,\Delta r^{SUSY})\nonumber\\
&=& 2|Q_f|{v_f\,a_f\over v_f^2+a_f^2}\left[({A\over \sqrt{1-A}}-4s^2)
({c^2\over s^2}\Delta\rho^{SUSY})+...\right]\,,
\nonumber\\
\nabla_{V_f}^{SUSY} &=&\delta\rho_{V_f}-8|Q_f|s^2\,{v_f\,a_f\over
v_f^2+a_f^2}\,
\delta\kappa_{V_f}
= 2 \left({v_f\delta v_f+a_f\delta a_f\over v_f^2+a_f^2}
\right)_{vertices}\,,
\label{eq:d123}
\end{eqnarray}
$\nabla_U$ being a universal correction, $\nabla_{Q_f}$ applies only to charged
fermions
and $\nabla_{V_f}$ is the vertex correction. For the first two types of
corrections, we
have singled out the $\Delta\rho^{SUSY}$ component entering the full
expressions, though
this does not mean that the rest of the
contributions are comparatively negligible, in contrast to what happens in
the SM case. As for the structure of $\nabla_{V_f}$, it is too complicated to
be discussed
here in any detail\,\cite{b1014}.

 It should be emphasized that
in the MSSM all potential quantum effects that
entail a departure with respect to the RSM are subdued by the decoupling
theorem.
This is because the breaking of SUSY is independent of the breaking of the
gauge symmetry. Therefore, in view of the current limits on sparticle masses
($m_{SUSY}\geq{\cal O}(M_Z) $),
the SUSY quantum effects are generally expected to be tiny as
compared to quantum effects from the RSM. This is indeed the case for the
radiative corrections to $M_W$\,\cite{b1002,b1005,b1006},
the reason being that, in contradistinction
to the light fermions in the SM, no leading RG-corrections
(first term on the RHS of eq.(\ref{eq:DeltarRG})) from sparticles
are possible in the MSSM. For the $Z$-width, however, this feature has a lesser
impact, since $\Delta\alpha$ cancels out in $\rho_f$\,\cite{b1009} (see later
on).
We are thus left with oblique corrections at the
next-to-leading  order (second and third
terms on the RHS of eq.(\ref{eq:DeltarRG})), plus non-oblique contributions
driven by significant Yukawa couplings. In both cases, the SM and SUSY effects
may be of the same order of magnitude, if sparticles are not too heavy.
Among the next-to-leading oblique corrections, we have the custodial symmetry
breaking ones. However, custodial symmetry cannot be broken in the
MSSM by non-decoupling SUSY effects, whether statical
($\Delta\rho$) or
dynamical (wave-function renormalization of the gauge bosons ).
For large enough $m_{SUSY}$, oblique and non-oblique SUSY corrections must go
to zero.
We have checked this analytically in our calculations, and also numerically in
our computer codes.
For example, consider the SUSY mass parameters
$(\mu, M)$ associated to the higgsino-gaugino parameter space\,\cite{b1001}.
In the mass-eigenstate basis, the 2-component
gaugino and higgsino Weyl spinors combine among themselves to form
4-component charginos and neutralinos (``inos'')
and the corresponding neutral
current is in general an admixture of vector and axial
components
(as is also the case for the neutral current associated
to conventional fermions).
However, for $M,\mu\rightarrow\infty$ the SUSY neutral current
becomes a pure vector-like current and
a standard vector Ward identity insures an exact cancellation of the
renormalized vertex functions constructed from ``inos'' and sfermions
in that limit. In short:
we expect to see measurable SUSY quantum effects on the $Z$-width only if the
soft SUSY-breaking mass parameters are not much larger than the electroweak
scale. In such circumstances, specific corrections
 (e.g. electroweak) from
SUSY can be comparable or even larger than in the SM, as it happens to be the
case e.g. for the
electroweak corrections to the top quark width\,\cite{b1B11} and also in the
present case for the $Z$-width. This will be demonstrated below.


We come now to explicit numerical results. The actual calculation
to obtain these results is rather
cumbersome, since we retain exact dependence on all masses and keep track of
all
matrix coupling constants for all SUSY particles in their respective
mass-eigenstate
bases\,\cite{b1014}. The computation of the many $3$-point functions involved
has been carried out using thoroughly tested  standard techniques
based on exact reduction formulae
and subsequent expansion of the scalar functions in terms of complex
Spence functions\,\cite{b3SPE}.
In Figs.1a and 1b we display contour lines of constant value of the genuine
SUSY
correction defined in eqs.(\ref{eq:Tshift}), (\ref{eq:DeltaSUSY})
\begin{equation}
 \delta\Gamma_Z^{SUSY} =
\sum_{l=e,\mu,\tau}\left[\delta\Gamma^{SUSY}(Z\rightarrow l^{+}l^{-})+
\delta\Gamma^{SUSY}(Z\rightarrow \nu_{l}\bar{\nu}_l)\right]
 +\sum_{q=u,d,c,s,b}\delta\Gamma^{SUSY}(Z\rightarrow q\bar{q})
\label{eq:GZSUSY}
\end{equation}
in a standard window of the higgsino-gaugino
$(\mu,M)$-space for Models I and II,
respectively. In both models, the pattern of sfermion masses is
generated from the generic formula\,\cite{b1001}
\begin{equation}
m^2_{{\tilde{f}}_{L,R}}=m^2_f+M^2_{{\tilde{f}}_{L,R}}\pm
\cos{2\beta}\,(T^3_{L,R}-Q_{\tilde{f}}\,\,s^2)\,M_Z^2\,,
\label{eq:sferm}
\end{equation}
where the same soft SUSY-breaking parameter
$M^2_{\tilde{f}_L}$ is shared by the two members of any $SU(2)_L$ doublet.
However, while in Model\, I the R-type sfermions are assumed to be degenerate
in mass
with the L-type ones, this is not so in
in Model\, II  where
one further specifies the structure of the soft SUSY-breaking mass
parameters in the standard manner suggested by radiative symmetry-breaking
models (such as supergravity inspired models\,\cite{b1001}), namely:
$M_{{\tilde{f}}_{L,R}}^2=m_0^2+C({\tilde{f}}_{L,R})\,M^2$, where $m_0$
is a universal soft SUSY-breaking
scalar mass at the GUT scale and $C({\tilde{f}}_{L,R})$ are
certain RG-driven coefficients\,\cite{b1IBLO}.
Altogether these models cover a sufficiently
wide range of phenomenologically allowed masses
for  sneutrinos
($\tilde{\nu}_l$), charged sleptons and squarks
($\tilde{l}^{\pm}_a, \tilde{q}_a\,(a=1,2)$),
charginos ($\Psi_i^{\pm}\,,i=1,2$) and
neutralinos ($\Psi_{\alpha}^{0}\,,\alpha=1,...,4)$, whose present
lower limits are\,\cite{b3001}
\begin{equation}
m_{\tilde{l}^{\pm}_a}\geq 45\,GeV\,,\ \ m_{\tilde{\nu}_l}\geq 42\,GeV
\,,\ \ m_{\tilde{q}_a}\geq 130\, GeV\,,
\label{eq:boundA1}
\end{equation}
\begin{equation}
 \ \ M_{\Psi^{\pm}_i}\geq 47\,GeV\,,  \ \ M_{\Psi^{0}_{\alpha}}\geq 20\,GeV\,.
\label{eq:boundA2}
\end{equation}
It should be pointed out that the squark mass limits from Tevatron are not
fully model
independent\,\cite{b3001} and in particular they depend on assumptions on the
gluino masses and on canonical SUSY decay modes. As a consequence, a stop
squark
$\tilde{t}_1$ could still be rather light
($\stackrel{\scriptstyle <}{{ }_{\sim}} M_Z/2$), a feature that can be easily
accomodated in model building
through a large mixing term $m_t M_{LR}$
(proportional to the top quark mass $m_t$)
in the stop mass matrix. In spite of the fact that we have not
included this term for the third squark family in eq.(\ref{eq:sferm}),
we shall amply exploit this possibility in Part II. Here
we prefer to present more conservative results by
keeping all squark generations alike, i.e. without mixing.
As in Ref. \cite{b1005},
we have furthermore imposed on our numerical analysis the condition
\begin{equation}
|\Delta\rho|^{SUSY}< 0.005\,.
\label{eq:rho}
\end{equation}
In Fig.1 we have fixed the
sfermion spectrum (\ref{eq:sferm}) with
$\tan{\beta}=8$, $m_{\tilde{\nu}_l}=50\,GeV$
and $m_{\tilde{u}}=m_{\tilde{c}}=m_{\tilde{b}}=130\,GeV$,
i.e. a spectrum perfectly consistently with the bounds (\ref{eq:boundA1}).
In particular, our conservative
choice for $m_{\tilde{b}}$ implies, via eq.(\ref{eq:sferm}) with $m_f=m_t$,
rather heavy ($\simeq 200\,GeV$) partners of the top quark.
It becomes patent from Fig.1a that genuine SUSY contributions of order
$\delta\Gamma_Z^{SUSY}\stackrel{\scriptstyle >}{{ }_{\sim}}+10\,MeV$
can comfortably be achieved in Model\, I, for a wide range of
chargino masses.
The fact that light charginos, i.e. charginos
corresponding to points $(M,\mu)$ near the boundary
of the phenomenologically allowed region in Fig.1a, are responsible for
a minimum SUSY correction to the $Z$-width is
related to the large (oblique) negative
contributions from the ``ino'' sector near
that boundary. (We shall take advantage of this
feature in Part\, II in connection to the analysis of $R_b$ in the MSSM).
 However, in the middle region
($M\simeq |\mu|\stackrel{\scriptstyle >}{{ }_{\sim}} 100-150\,GeV$),
 the latter contributions are positive (though smaller in absolute value) and
add up to
the oblique sfermion contributions plus the
non-oblique (vertex) corrections, which are also globally positive in this
case, and
consequently
$\delta\Gamma_Z^{SUSY}$ increases up to a maximum $\simeq +13\,MeV$.
 Of course, away from the local maximum, the total SUSY contribution diminishes
for larger and larger values of $(\mu,M)$, but it does not go to zero; instead,
$\delta\Gamma_Z^{SUSY}$ tends asymptotically to the constant
positive effect from the sfermion self-energies,
which have fixed values for their masses,
and it would only decouple upon simultaneous increase of the latters.
 It is worthwhile to note that sufficiently
away from the neighbourhood of the boundary
(specifically, for $M,|\mu|>150\,GeV$)
the SUSY quantum effects described in Fig.1a originate
from an average SUSY spectrum
which is beyond the capability of pair production at LEP 200.
Indeed, given the values of the sfermion masses fixed above, in this region we
have
\begin{equation}
m_{\tilde{l}^{\pm}_a}\geq 94\,GeV\,, \ \ m_{\tilde{q}_a}\geq 130\, GeV\,,
\ \ M_{\Psi^{\pm}_i}\geq 94 \,GeV\,,  \ \ M_{\Psi^{0}_{\alpha}}\geq 55\,GeV\,.
\label{eq:boundB}
\end{equation}
Remarkably enough, it turns out that even in this LEP 200 unaccessible region,
the quantum correction
$\delta\Gamma_Z^{SUSY}$ may well reach the level of the total error
(theoretical plus experimental, added in quadrature)
in eqs.(\ref{eq:GZexp},\ref{eq:GZMSM1}-\ref{eq:GZMSM2})
and it therefore leaves open the possibility to potentially detect these
extra effects in the future. It follows that in most of the window of Fig.1a,
the average electroweak
SUSY correction to the $Z$-width is numerically larger,
but opposite in sign, as compared to the (negative) electroweak SM correction
$\delta\Gamma_Z^{\rm (ew) SM}=\Gamma_Z^{\rm (ew) SM}-{\Gamma}_Z^0(G_F)$
with respect to the SM tree-level width in the
$G_F$-parametrization.
We have checked this explicitly using the upgraded version of the
computer code BHM\,\cite{b3BHM}, with the inputs:
$M_Z$ from eq.(\ref{eq:MZexp}),
$m_t=174\,GeV$, $M_H=100\,GeV$
\footnote{Since we are assimilating the SM to the RSM,
the lightest neutral Higgs mass must
be less than about $130\,GeV$\,\cite{b1313}.}
and then projecting the electroweak part, with the result
$\delta\Gamma_Z^{\rm (ew) SM}=-9.5\,MeV$.
Therefore, if the physical width of the $Z$ turns out to be the one predicted
by
the MSSM, eq.(\ref{eq:GZMSSM}), the total electroweak correction
with respect to the tree-level width $\Gamma_Z^0(G_F)$ in the SM (now the RSM)
will be (neglecting for the moment the extra Higgs correction)
\begin{equation}
 \delta\Gamma_Z^{\rm (ew) MSSM}=
\Gamma_Z^{\rm (ew) MSSM}-{\Gamma}_Z^0(G_F)=
\delta\Gamma_Z^{\rm (ew) SM}+\delta\Gamma_Z^{SUSY}
\simeq +(1-3) \,MeV.
\end{equation}
 This will hold true for a SUSY spectrum satisfying eq.(\ref{eq:boundB}).
For demonstration purposes we have tolerated an overall SUSY correction
slightly
exceeding the total theoretical plus experimental
error from eqs.(\ref{eq:GZexp},\ref{eq:GZMSM1}-\ref{eq:GZMSM2}), just to
exhibite the
potentiality of the supersymmetric
 virtual effects from a sparticle spectrum amply complying
with the present phenomenological bounds.
For heavier and heavier sparticles, the
correction decreases and can be made numerically very
close to the electroweak correction,
but opposite in sign. The resulting cancellation could
effectively render invisible the total electroweak correction in the MSSM,
even assuming a substantial improvement of the theoretical errors.
We conclude that the effect might perhaps
be discovered either by ``missing'' the expected electroweak correction in the
SM, or even finding that it goes in the opposite direction.
 Of course, as we have advertised,  this conclusion
would only apply if the negative Higgs effects in the MSSM
($\delta\Gamma_Z^H$ in eq.(\ref{eq:Tshift}))
 are kept very small (in absolute value)
 with respect to $\delta\Gamma^{SUSY}_Z$, or if we confine our study into a
region
where the additional Higgs effects from the MSSM are always positive. Indeed,
as
shown in Part\, II, we are naturally led to a particular scenario like this
when trying to
cope with the ``$R_b$ crisis'' in the MSSM.
Notwithstanding, it should be mentioned that
global negative corrections $\delta\Gamma_Z^{MSSM}$
are still possible by picking contrived values for the parameters.

The corresponding contour lines for Model\, II are shown
in Fig.1b. In this case the sfermion masses are controlled by the single
parameter $m_0$.
Since a minimum value
common for squarks and sleptons must be found out for this parameter
($m_0\stackrel{\scriptstyle >}{{ }_{\sim}}60\,GeV$)
in order to simultaneously fulfil all the phenomenological bounds
(\ref{eq:boundA1}),(\ref{eq:boundA2}),
the average sfermion spectrum in Model\, II turns out to be far
more heavier than in Model\, I. For example, in the middle region of Fig.1b
($M,|\mu|\geq 150\,GeV$) the sfermion spectrum in Model\, II satisfies
\begin{equation}
m_{\tilde{l}^{\pm}}\,, m_{\tilde{\nu}_l}> 100\,GeV
\,,\ \ m_{\tilde{q}}> 400\, GeV\,.
\label{eq:boundC}
\end{equation}
Therefore, the corrections are generally smaller (about a factor of $2-3$) than
in Model\, I,
though still not fully negligible: around the maximum,
 $\delta\Gamma_Z^{SUSY}\simeq +6\,MeV$,
i.e. numerically close to
the SM electroweak correction\,\cite{b1009,b3BHM},
and reversed in sign.
Notice that there are no threshold effects neither in Model\, I nor in
Model\, II associated to
the wave function renormalization of
the $Z$ gauge boson, eq.(\ref{eq:RSEPZ}). In fact, while sfermion
and chargino masses are always assumed to be
heavier than half the $Z$-mass, the neutralino masses may continuously approach
points of $(\mu,M)$-space where $M^0_{\alpha}+M^0_{\beta}=M_Z$. However, the
corresponding singularity in $\hat{\Sigma}_Z'(M_Z^2)$ cancels out identically
and
no threshold effect remains.

For any channel $Z\rightarrow f\bar{f}$,
$\nabla_{U,Q}^{SUSY}$ from eq.(\ref{eq:d123})
give the leading
contribution, which is numerically very close for the two $T^3=\pm 1/2$
components in
each fermion doublet. Therefore,
the vertex correction, $\nabla_{V_f}^{SUSY}$, which is negative for
the up components and positive for the down components,
gives the bulk of the difference
between the corrections to the $Z$ partial widths into
$T^3= +1/2$ fermions and $T^3=- 1/2$ fermions (cf. Figs.2-5).
It is worth mentioning that although $\nabla_{U,Q}^{SUSY}$ are leading,
there are strong cancellations between the two terms in these formulas; for
example,
as seen on the first eq.(\ref{eq:d123}), the $\Delta\alpha$
contribution from eq.(\ref{eq:DeltarRG}) exactly cancels in the difference
 $\nabla_U^{SUSY}$,
leaving $\Delta\rho^{SUSY}$ as one of the leading
remainders. On the other hand,  the vertex
correction is specially significant for the $b\bar{b}$ channel, where the
Yukawa couplings
can be substantially large.
Nevertheless, in contrast to the SM, where there is an
overcompensation of the (positive) propagator correction by the
(negative) vertex correction\,\cite{b1009}, it turns out that
the extra non-oblique SUSY contribution is of the same
order of magnitude and has the same (positive)
sign as the extra oblique contribution from $\Delta\rho^{SUSY}$. As a
consequence, the
two effects turn out to add up and might give
rise to a measurable correction. We show this explicitly
in Figs.2a-2d, where we fix $(\mu,M)=(-100,100)\,GeV$ and plot the dependence
of the partial width corrections, $\delta\Gamma_Z^{SUSY}(Z\rightarrow
f\bar{f})$,
on the squark and slepton masses in
Model\, I. Upon inspection of these figures
we gather three noticeable facts:
 i) the SUSY corrections to the partial widths
are about $4-8$ times
larger for quarks than for leptons, ii) the corrections to the partial widths
into $T^3=-1/2$ fermions are larger than the corrections to the partial widths
into
$T^3=+1/2$ fermions, the
 $b\bar{b}$ channel being the most
favoured one,
and iii) there is a moderate decrease of the corrections for heavier and
heavier
sfermions masses;
e.g. in Fig.2a, $\delta\Gamma^{SUSY}(Z\rightarrow b\bar{b})$ roughly
decreases from $4\,MeV$ to $2.7\,MeV$ when
$m_{\tilde u}$ increases from $100\,GeV$
to $200\,GeV$. The decrease is even slower in Fig.2c, where the partial widths
into quarks are confronted with increasing slepton masses, which enter through
oblique
corrections the weak gauge boson propagators. Similar considerations apply,
respectively, to Figs.2b and 2d.
The corresponding plots for Model\, II are shown
in Figs.3a-3b. In this case the
relevant parameter is the universal soft SUSY-breaking sfermion mass
$m_0$. Notice that some of the corrections may be slightly negative, as is the
case for the neutrino channels $\nu_{l}\bar{\nu}_l$ in Fig.3b.

In Figs.4a-4b, we study the evolution of the radiative corrections on the
parameter
$\tan{\beta}$, only for Model\, I, where the effects are bigger.
The partial widths into leptons and quarks are
separately considered. There
we see why we have chosen $\tan{\beta}=8$ in Fig.1; it corresponds to a
an approximate saturation value
for the leading ($q\bar{q}$) contributions
in the range $\tan\beta\stackrel{\scriptstyle <}{{ }_{\sim}} 50$.
For the leptons, however, especially for the $\tau^{+}\tau^{-}$ and
$\nu_{\tau}\bar{\nu}_{\tau}$ channels, there is still
an ongoing evolution beyond $\tan{\beta}\geq 8$,
though the relative impact on $\delta\Gamma_Z^{SUSY}$
is minor. Remember that $\tan{\beta}$ enters
the mass matrices for chargino-neutralinos and also determines
potentially relevant fermion-sfermion-chargino/neutralino Yukawa couplings
in the vertex corrections $\nabla_{V_f}^{SUSY}$ on eq.(\ref{eq:d123}).
The most significant ones are
\begin{equation}
h_t={g\,m_t\over \sqrt{2}\,M_W\,\sin{\beta}}\;\;\;\;\;,
\;\;\;\;\; h_b={g\,m_b\over \sqrt{2}\,M_W\,\cos{\beta}}\,,
\label{eq:Yukawas1}
\end{equation}
which contribute to the SUSY effects on the $Zb\bar{b}$ vertex, and
\begin{equation}
h_{\tau}={g\,m_{\tau}\over \sqrt{2}\,M_W\,\cos{\beta}}\,,
\label{eq:Yukawas2}
\end{equation}
which goes into the structure of the $Z\tau^{+}\tau^{-}$
and $Z\nu_{\tau}\bar{\nu}_{\tau}$
vertices. The Yukawa coupling
associated to the charm quark, $h_c$,  is not sizeable enough since $m_c/M_W$
is
too small and $h_c$ does not increase with $\tan{\beta}$. That is why the
contributions
to the partial widths into the modes
 $u\bar{u}$ and $c\bar{c}$ are practically indistinguishable
in our figures.
Obviously, another point where to focus our attention is on the dependence
of $\delta\Gamma_Z^{SUSY}$ on the top quark mass (Figs.5a-5b). Since this
parameter
is involved in the sbottom-stop mass differences,  all
partial widths
are universally affected (through propagator corrections)
 by the precise value of $m_t$, and
$\delta\Gamma^{SUSY}(Z\rightarrow b\bar{b})$
is additionally affected by important vertex corrections
associated to an enhancement of $h_t$ for increasing $m_t$.
Indeed, the slope of the $b\bar{b}$ curve can be
seen from Fig.5a to be
more pronounced than for the other channels. As a matter of fact,
of the three Yukawa couplings (\ref{eq:Yukawas1})-(\ref{eq:Yukawas2}), the
most significant one is $h_t$. It basically determines the important
non-oblique
SUSY corrections responsible for the gap between
the $b\bar{b}$ channel with respect to the the $s\bar{s}$ and
$d\bar{d}$ modes in Figs.2-5. For large enough $\tan{\beta}$, the
corrections to the $b\bar{b}$ channel are further enhanced by an increasing
$h_b$ (with $m_b=5\,GeV$),
as can be appreciated in Fig.4a. To be more specific, there is a balance
between
chargino and neutralino contributions in the vertex corrections; namely,
the chargino vertex corrections
diminish with $\tan\beta$ whereas the neutralino vertex corrections increase
with it.
Only for large enough $\tan{\beta}$ the neutralino effects are overwhelming;
for
example, for
$\tan{\beta}\stackrel{\scriptstyle >}{{ }_{\sim}} 70$
(where we are bordering the limits of validity of
perturbation theory) and keeping the same values for the rest of the
parameters,
$\delta\Gamma^{SUSY}(Z\rightarrow b\bar{b})\stackrel{\scriptstyle >}{{
}_{\sim}} 4\,MeV$.
 On the other hand, even though we wish not to emphasize the
region of $\tan{\beta}< 1$, it is worthwhile to note that, there,
$h_t$ grows so fast that one could easily obtain similar staggering effects as
before,
viz. $\delta\Gamma^{SUSY}(Z\rightarrow b\bar{b})
\stackrel{\scriptstyle >}{{ }_{\sim}} 4\,MeV$ for
$\tan{\beta}\stackrel{\scriptstyle <}{{ }_{\sim}} 0.7$
\footnote {Scenarios with $\tan{\beta}<1$ are generally
disfavoured\,\cite{b3002} but
still some reduced interval
$0.7\stackrel{\scriptstyle <}{{ }_{\sim}}\tan{\beta}< 1$ is in dispute.}.
In this region, which we have not
highlighted in Fig.4, there is in fact an unbounded, positive, contribution
at the left edge of the curves. Nevertheless
we shall see in Part\, II that this contribution
is compensated by an overwhelming negative effect from
$\delta\Gamma_Z^{H}$. We remark that the point
 $\tan{\beta}=1$ is the site of the absolute minimum for each of the partial
widths,
since at this point
the sfermion generations have minimal mass splittings (cf.eq.(\ref{eq:sferm}))
and hence $\Delta\rho^{SUSY}$ takes on its smallest
value.

In conclusion, there could be significant, genuinely supersymmetric,
electroweak
renormalization effects on the $Z$-width in the MSSM. We have explored a
sufficiently
wide domain of the SUSY parameter space in support of this fact.
In particular, we have concentrated
our numerical analysis on regions
corresponding to rather conservative values for all sparticle masses; thus,
although it is not strictly required by the phenomenological bounds,
we have assumed masses of ${\cal O}(100)\,GeV$ for the superpartners of the top
quark, and in general
we have put more emphasis on the results obtained
for a full SUSY spectrum above the possibilities of
pair production at LEP 200. Even with these hypotheses,
the size of the ensuing SUSY corrections
could easily reach the level of the present theoretical errors.
For the recent CDF values of the top quark mass, we have found
natural windows of parameter space where there could be a large cancellation,
or even an overcompensation,
of the expected electroweak SM corrections by the electroweak supersymmetric
quantum
effects.
Upon relaxing the previous hypotheses on sparticle masses
to the strict limit placed by the present
LEP 100 bounds, more freedom is obtained to test the
potentiality of the SUSY corrections in $Z$-decay physics. This point will be
retaken and fully exploited in Part\, II.
We hope that in the future a better determination of $m_t$, $\alpha_s$
and the parameters of the Higgs sector, will allow
to uncover the electroweak quantum effects on  $\Gamma_Z$. Missing of these
effects, or finding them opposite in sign to SM expectations,
could be interpreted as indirect evidence of SUSY.

\vspace{0.5cm}

{\bf Acknowledgements}: One of us (JS) is thankful to Wolfgang Hollik for
useful
discussions and gratefully acknowledges
the hospitality at the Institut f\"ur Theoretische Physik
der Universit\"at Karlsruhe during a visit.
He also thanks P. Chankowski and A. Dabelstein for discussions and for
informing him
on similar calculations being carried out in that Institut.
We are grateful to M. Mart\'{\i}nez and F. Teubert for helping us in the use
of the computer code BHM.
Finally, we
are indebted to A. Pascual and F. S\'anchez for their assistance in the
technical
performance of the figures using PAW.
This work has  been partially supported by CICYT
under project No. AEN93-0474. The work of DG has also been supported by a grant
of
the Comissionat per a Universitats i Recerca, Generalitat de Catalunya.

\vspace{2.25cm}
\begin{center}
\begin{Large}
{\bf Figure Captions}
\end{Large}
\end{center}
\begin{itemize}
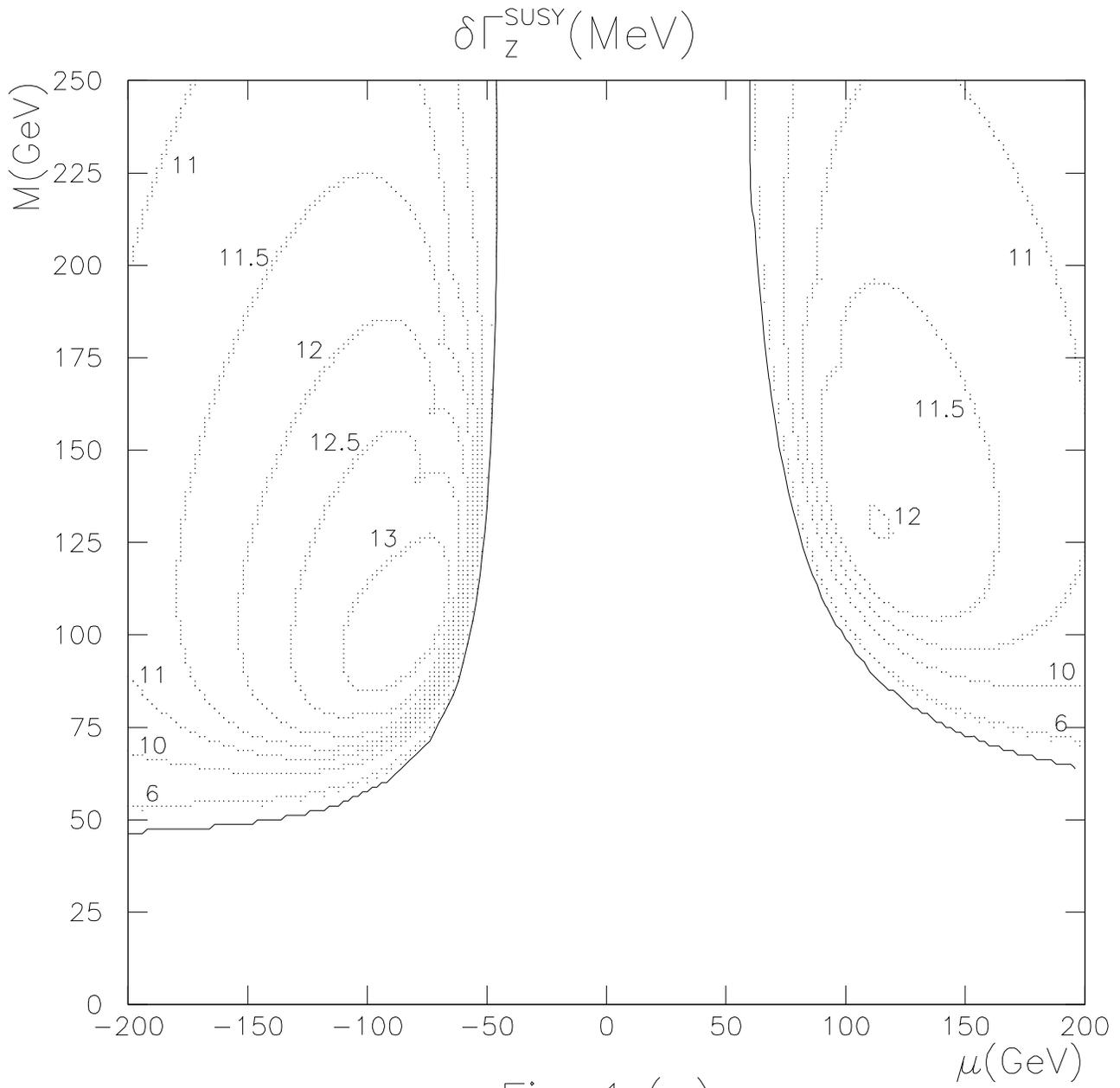
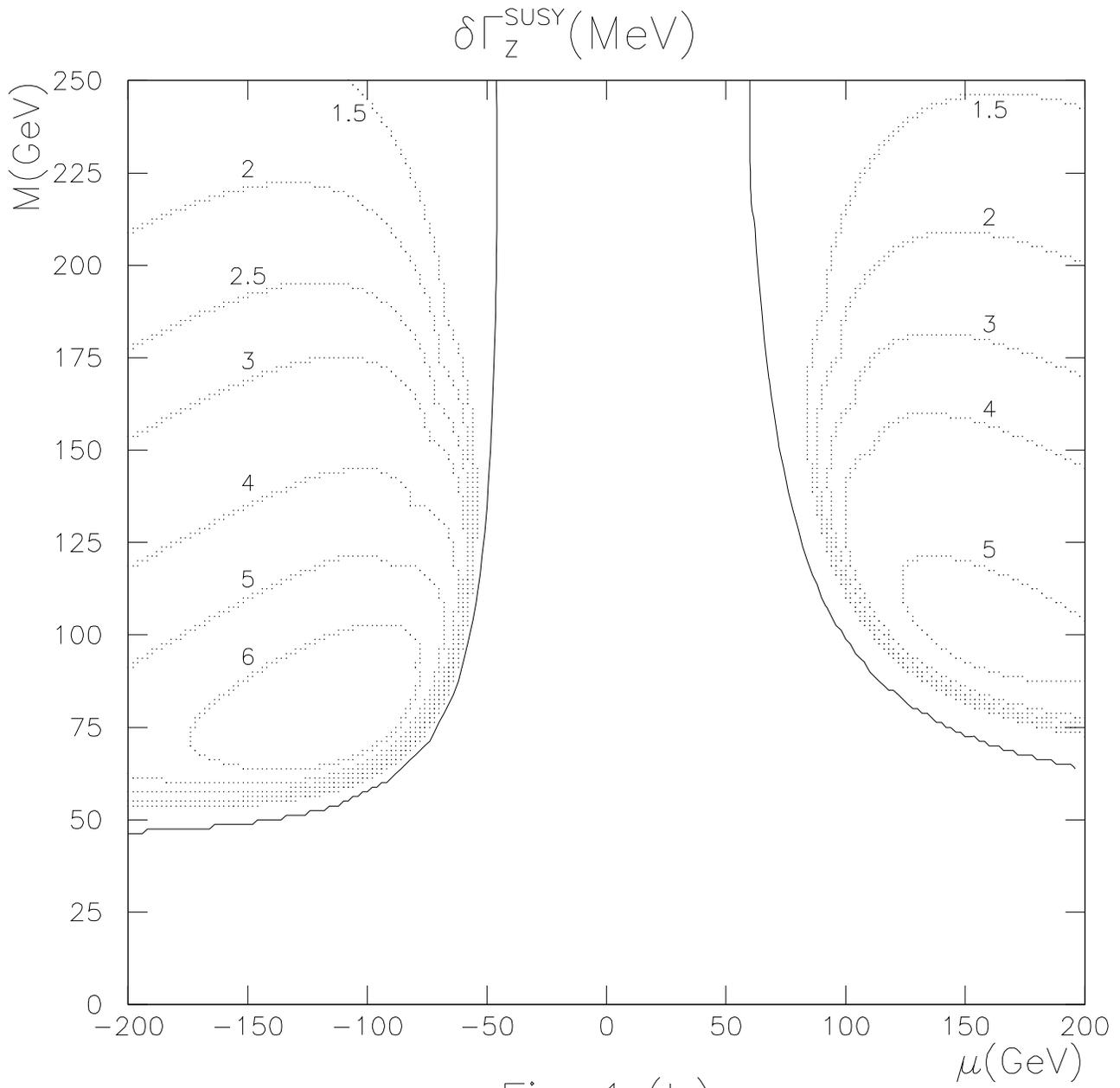
\item{\bf Fig.1} (a) Contour plots in the higgsino-gaugino $(\mu, M)$-parameter
space for the total SUSY correction $\delta\Gamma_Z^{SUSY}$ in Model\, I.
The sfermion spectrum is obtained from eq.(\ref{eq:sferm})  with
$\tan{\beta}=8$, $m_{\tilde{\nu}_l}=50\,GeV$ and
$m_{\tilde{u}}=m_{\tilde{c}}=m_{\tilde{b}}=130\,GeV$.
The top quark mass is fixed at $m_t=174\,GeV$.
The blank regions are phenomenologically excluded
by the constraints
$M_{\Psi^{\pm}_i}> 47\,GeV$, $M_{\Psi^{0}_{\alpha}}> 20\,GeV$;
(b) As in case (a), but for Model\, II and fixed $m_0=63\,GeV$.

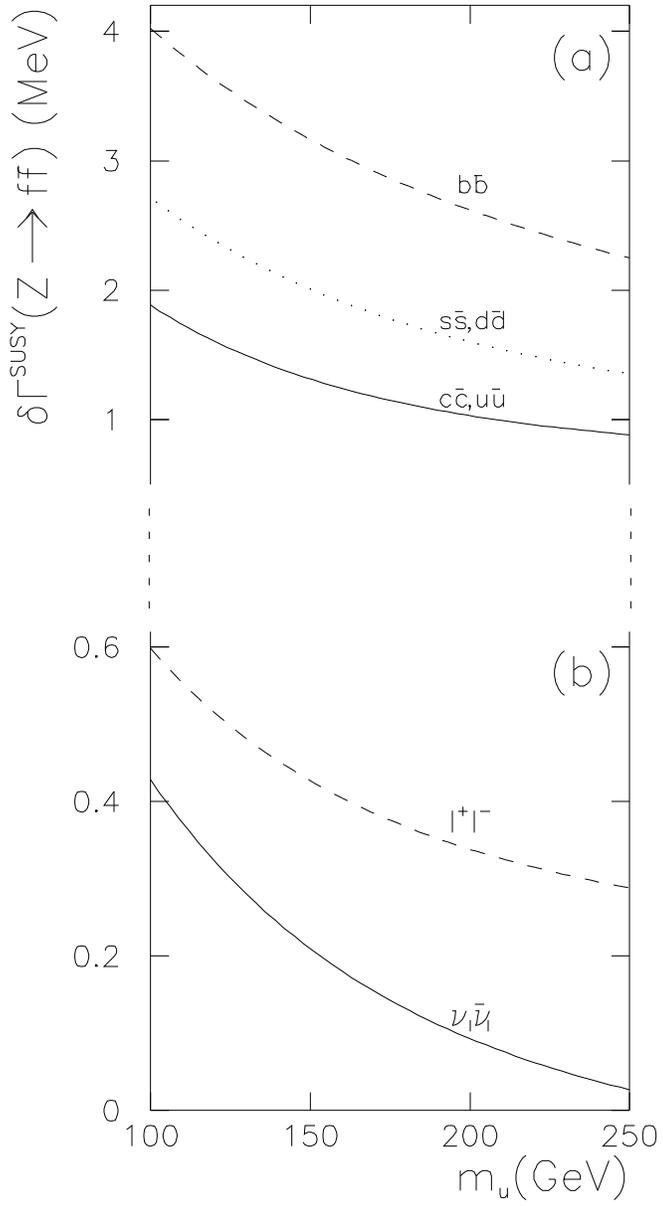
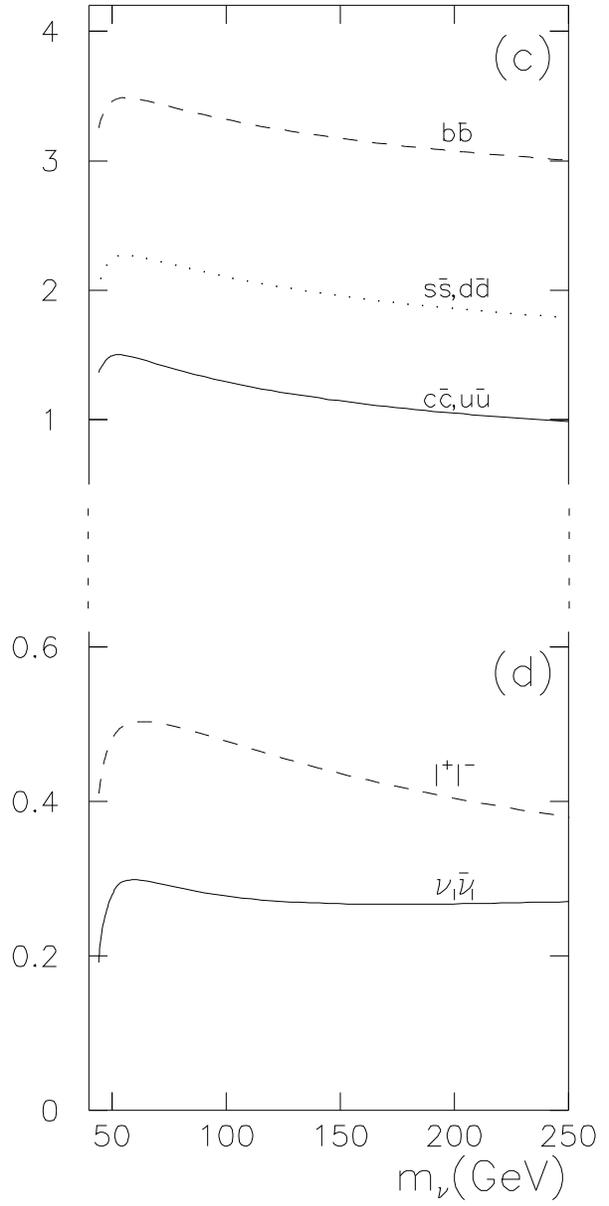
\item{\bf Fig.2} SUSY corrections to the various partial widths
 $\Gamma (Z\rightarrow f\bar{f})$ in Model\, I as a function of the
squark masses (cases (a) and (b)) and slepton masses (cases (c) and (d)).
We have taken $(\mu, M)=(-100,100)\,GeV$ and the other non-varying
parameters as in Fig. 1a.

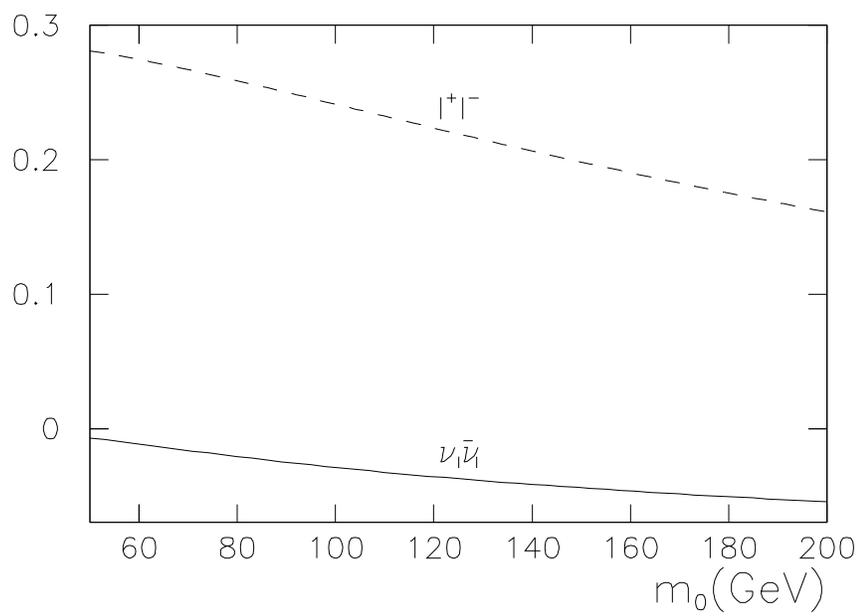
\item{\bf Fig.3} As in Fig.2, but for Model\, II as a function
of $m_0$.

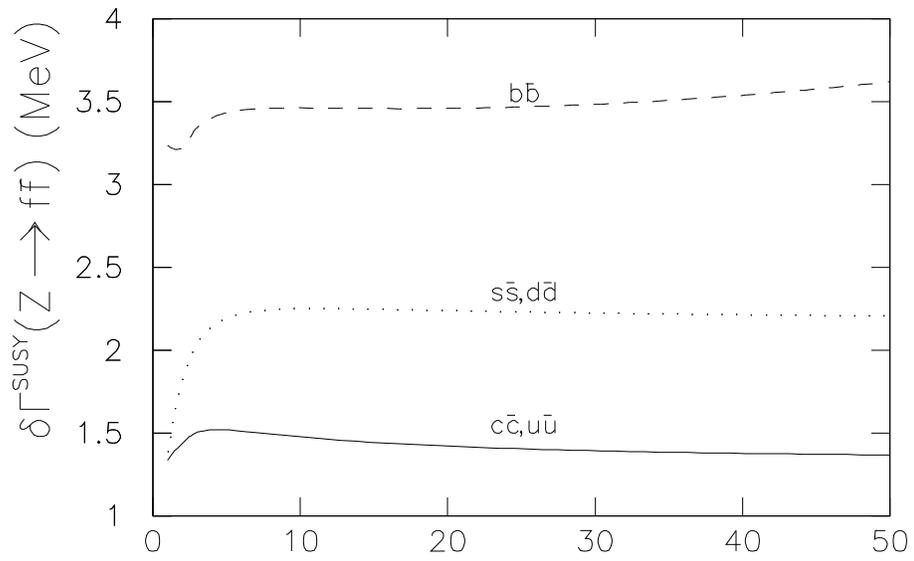
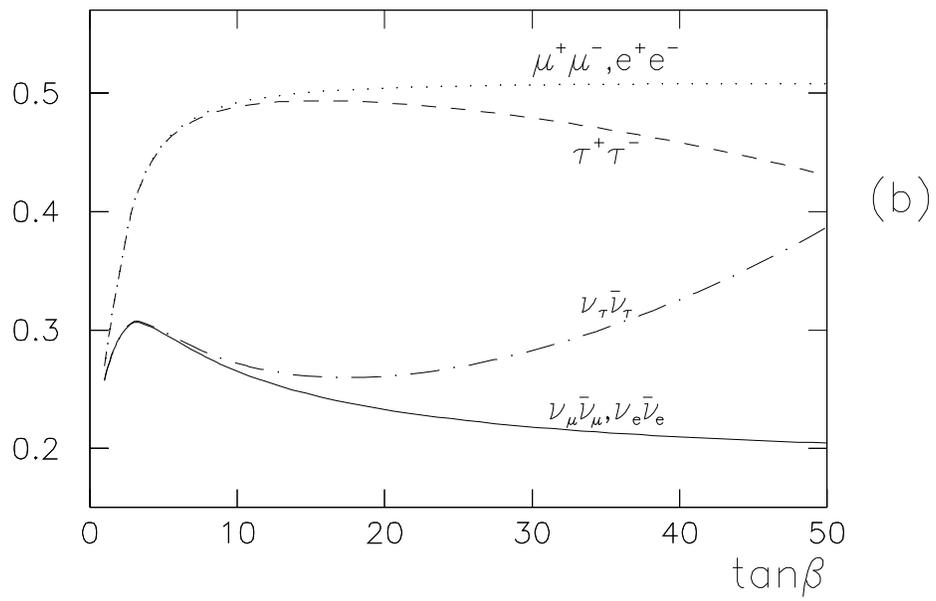
\item{\bf Fig.4} Dependence of $\delta\Gamma^{SUSY}(Z\rightarrow f\bar{f})$
on $\tan{\beta}$ for $Z$ decaying into (a) quark-antiquark, and (b)
lepton-antilepton in Model\, I. The remaining fixed
parameters are chosen as in Fig.1a and Fig.2.

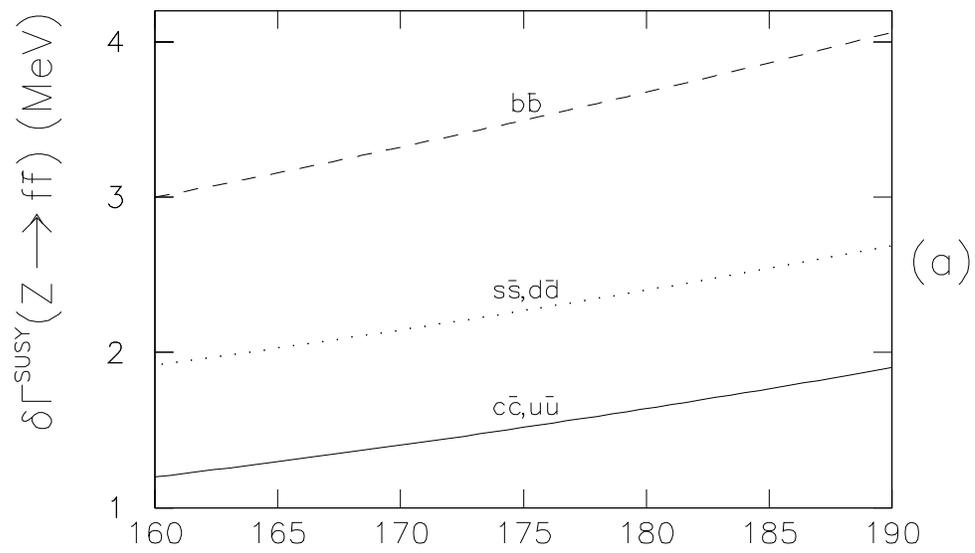
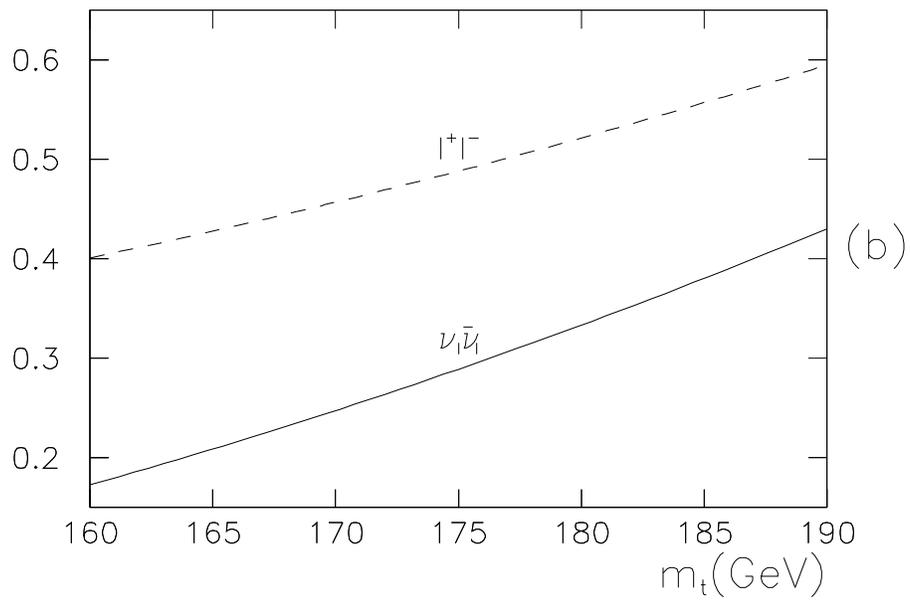
\item{\bf Fig.5} Dependence of $\delta\Gamma^{SUSY}(Z\rightarrow f\bar{f})$
on $m_t$ within the CDF mass limits,
for $Z$ decaying into (a) quark-antiquark, and (b)
lepton-antilepton in Model\, I. Remaining parameters as in Fig.1a and Fig.2.

\end{itemize}


\vspace{1.75cm}
\begin{thebibliography}{99999}



\bibitem{b1001}
H. Nilles, Phys. Rep. {\bf 110} (1984) 1; H. Haber and G. Kane, Phys. Rep.
{\bf 117} (1985) 75;
 A. Lahanas and D. Nanopoulos, Phys. Rep. {\bf 145} (1987) 1;
 See also the exhaustive reprint collection {\bf Supersymmetry}
(2 vols.), ed. S. Ferrara (North Holland/World Scientific, Singapore, 1987);
H.E. Haber, {\bf Introductory Low-Energy Supersymmetry}, preprint SCIPP-92/33,
Proc. of the 1992 TASI, Univ. of Colorado.
\bibitem{b1002}
R. Barbieri, M. Frigeni,
F. Giuliani and H. Haber, Nucl. Phys. {\bf B341} (1990) 309; R. Barbieri,
M. Frigeni and F. Caravaglios, Phys. Lett.{\bf B279} (1992) 169;
G. Altarelli, R. Barbieri and F. Caravaglios, Phys. Lett. {\bf B324} (1993)
357.
\bibitem{b1003}
See e.g. M. Consoli, W. Hollik, G. Burgers and F. Jegerlehner,
in: {\bf Z physics at LEP 1}, eds. G. Altarelli, R. Kleiss
and C. Verzegnassi, Yellow Report CERN 89-08 (1989).
\bibitem{b1004}
C.H. LLewellyn Smith, talk at the Institut de F\'{\i}sica d'Altes Energies
(IFAE), Univ. Aut\`onoma de Barcelona, 1993;
L. Camillieri et al, LEP Committee Report on the measurement of $M_W$ at
LEP II, November 1992.
\bibitem{b1005}
D. Garcia and J. Sol\`a, Mod. Phys. Lett. {\bf A9} (1994) 211.
\bibitem{b1006}
P.H. Chankowski, A. Dabelstein, W. Hollik, W. M\"osle, S. Pokorski and
J. Rosiek, Nucl. Phys. {\bf B417} (1994) 101.
\bibitem{b1007}
M. Pepe-Altarelli, talk at the Tennessee International Symposium on
Radiative Corrections, Gatlinburg, Tennessee, June 27-July 1, 1994 (to appear
in the Proceedings).
\bibitem{b10B8}
D.A. Ross and J.C. Taylor, Nucl. Phys. {\bf B51} (1973) 25;
 A. Sirlin, Phys. Rev.
{\bf D22} (1980) 971;
 W. Marciano and A. Sirlin, Phys. Rev. {\bf D22} (1980) 2695;
{\sl ibid} {\bf D29} (1984) 75 and 945.
\bibitem{b1009}
W. Beenaker and W. Hollik, Z. Phys. {\bf C40} (1988) 141
\bibitem{b1010}
M. Consoli, S. LoPresti and  L. Maiani, Nucl. Phys. {\bf B223} (1983) 474;
P. Antonelli, M. Consoli and C. Corbo, Phys. Lett. {\bf B99} (1981) 475;
F. Jegerlehner, Z. Phys. {\bf C32} (1986) 425; W. Wetzel, Nucl. Phys.
{\bf B227} (1983) 1; A.A. Akhundov, D. Yu. Bardin and T. Riemann, Nucl. Phys.
{\bf B276} (1986) 1; J. Bernabeu, A. Pich and A. Santamaria, Phys. Lett.
{\bf B200} (1988) 569.
\bibitem{b1011}
W. Hollik, talk at the Tennessee International Symposium on
Radiative Corrections, Gatlinburg, Tennessee, June 27-July 1, 1994 (to appear
in
the Proceedings).
\bibitem{b1TOP}
F. Abe et al. (CDF Collab.), Phys. Rev. Lett. {\bf 73} (1994) 225.
\bibitem{b1HAB}
H.E. Haber, in ref.\cite{b1001}.
\bibitem{b1B11}
D. Garcia, W. Hollik, R.A. Jim\'enez and J. Sol\`a,
Nucl. Phys. {\bf B427} (1994) 53.
\bibitem{b1012}
G. Girardi, W. Hollik and C. Verzegnassi, Phys. Lett. {\bf B240} (1990) 492;
A. Djouadi, G. Girardi, C. Verzegnassi, W. Hollik and F.M. Renard, Nucl. Phys.
{\bf B349} (1991) 48;
M. Boulware and D. Finnell, Phys. Rev. {\bf D44} (1991) 2054.
\bibitem{b1013}
J.F. Gunion, H.E. Haber, G. Kane and S. Dawson, {\bf The Higgs hunter's guide}
(Addison-Wesley, New York, 1990);
J.F. Gunion and H.E. Haber, Nucl. Phys. {\bf B272} (1986) 1; {\sl ibid} {\bf
B278}
(1986) 449;
W. Hollik, Z. Phys. {\bf C32} (1986) 291; {\sl ibid} {\bf C37} (1988) 569;
S. Bertolini, Nucl. Phys. {\bf B272} (1986) 77;
W. Hollik, Mod. Phys. Lett. {\bf A5} (1990) 1909; H.E. Haber, in
ref.\cite{b1001}.
\bibitem{b1113}
A. Djouadi, J.L. Kneur and G. Moultaka, Phys. Lett. {\bf B242} (1990) 265.
\bibitem{b1213}
 A. Denner, R.J. Guth, W. Hollik and J.H. K\"uhn, Z. Phys. {\bf C51} (1991)
695;
W. Hollik, Mod. Phys. Lett. {\bf A5} (1990) 1909.
\bibitem{b1QCD}
A. Djouadi, M. Drees and H. K\"onig, Phys. Rev. {\bf D48} (1993) 3081.
\bibitem{b1PII}
D. Garcia, R.A. Jim\'enez and J. Sol\`a, {\bf Supersymmetric electroweak
renormalization
of the $Z$-width in the MSSM (II)}, preprint UAB-FT-344 ( September, 1994).
\bibitem{b1313}
H.E. Haber and R. Hempfling, Phys. Rev. Lett. {\bf 66} (1991) 1815;
Y. Okada, M. Yamaguchi and T. Yanagida, Prog. Theor. Phys. {\bf 85} (1991) 1;
J. Ellis, G. Ridolfi and F. Zwirner, Phys. Lett. {\bf B257} (1991) 83;
R. Hempfling and A. Hoang, Phys. Lett. {\bf B331} (1994) 99.
\bibitem{b1014}
D. Garcia, R.A. Jim\'enez and J. Sol\`a, {\bf The width of the $Z$ boson in the
MSSM}, preprint UAB-FT in preparation; D. Garcia, PhD Thesis, in preparation,
Univ. Aut\`onoma de Barcelona; R.A. Jim\'enez, PhD Thesis, in preparation,
Univ.
Aut\`onoma de Barcelona.



\bibitem{b2001}
M. B\"ohm, H. Spiesberger and W. Hollik, Fortschr. Phys. {\bf 34} (1986) 687;
W. Hollik,  Fortschr. Phys. {\bf 38} (1990) 165.
\bibitem{b2002}
A. Denner,  Fortschr. Phys. {\bf 41} (1993) 307;
F. Jegerlehner, {\bf Renormalizing the Standard Model}, Proc.of the 1990 TASI,
Univ. of Colorado.
\bibitem{b2022}
M. Einhorn, D. Jones and M. Veltman, Nucl. Phys. {\bf B191} (1981) 146.
\bibitem{b2032}
M.E. Peskin and T. Takeuchi, Phys. Rev. Lett. {\bf 65} (1990) 964;
D.C. Kennedy, Phys. Lett. {\bf B268} (1991) 86.
\bibitem{b2003}
J. Grifols and J. Sol\`a, Nucl. Phys. {\bf B253} (1985) 47;
Phys. Lett. {\bf B137} (1984) 257; J. Sol\`a, in: {\bf Phenomenological
Aspects of Supersymmetry}, ed. W. Hollik, R. R\"uckl and J. Wess (Springer
-Verlag, Lecture Notes in Physics {\bf 405}, 1992).

%
\bibitem{b3SPE}
G. `t Hooft and M. Veltman, Nucl. Phys. {\bf B153} (1979) 365;
G. Passarino and M. Veltman, Nucl.Phys. {\bf B160} (1979) 151;
A. Axelrod, Nucl. Phys. {\bf B209} (1982) 349.
\bibitem{b1IBLO}
K. Inoue, A. Kakuto, H. Komatsu and S. Takeshita, Prog. Theor. Phys. {\bf 67}
(1982) 1889; {\sl ibid} {\bf 68} (1982) 927;
L.J. Hall and J. Polchinski, Phys. Lett. {\bf B152} (1985) 335;
L. Iba\~{n}ez, C. L\'opez and C. Mu\~noz, Nucl. Phys. {\bf B256} (1985) 218;
S. Bertolini, F. Borzumati, A. Masiero and G. Ridolfi, Nucl. Phys. {\bf B353}
(1991) 591;
M. Olechowski and S. Pokorski, Nucl. Phys. {\bf B404} (1993) 590.
\bibitem{b3001}
{\bf Ten Years of SUSY Confronting Experiment}, ed. J. Ellis, D.V. Nanopoulos
and A. Savoy-Navarro, preprint CERN-TH.6707/92-PPE/92-180.
\bibitem{b3BHM}
Computer code by G. Burgers, W. Hollik and M. Mart\'{\i}nez; M. Consoli,
W. Hollik and F. Jegerlehner: Proc. of the Workshop on $Z$ Physics at LEP1,
CERN 89-90, Sept. 1989, ed. G. Altarelli et al., Vol.1, p.7; G. Burgers,
F. Jegerlehner, B. Kniehl and J. K\"uhn: the same Proc. Vol.1, p.55.
\bibitem{b3002}
G.F. Giudice and G. Ridolfi, Z. Phys. {\bf C41} (1988)447;
M. Olechowski and S. Pokorski, Phys. Lett. {\bf B214} (1988) 393;
M. Drees and M.M. Nojiri, Nucl. Phys. {\bf B369} (1992) 54.


\end{thebibliography}
\end{document}